\documentclass[twocolumn,aps,prl,superscriptaddress]{revtex4-2}

\usepackage{amsmath}
\usepackage{graphicx}
\usepackage{MnSymbol}
\usepackage{parskip}
\usepackage{makecell}
\usepackage{booktabs}
\usepackage{multirow}

\begin{document}

\title{Flash temperature in sliding contacts: comparing theory with experiments}

\author{B.N.J. Persson}
\affiliation{Peter Gr\"unberg Institute (PGI-1), Forschungszentrum J\"ulich, 52425, J\"ulich, Germany}
\affiliation{State Key Laboratory of Solid Lubrication, Lanzhou Institute of Chemical Physics, Chinese Academy of Sciences, 730000 Lanzhou, China}
\affiliation{MultiscaleConsulting, Wolfshovener str. 2, 52428 J\"ulich, Germany}

\begin{abstract}
The temperature increase in the contact regions between solids in sliding contact
has a huge influence on friction and wear. Here we test an analytical 
theory for the flash temperature, valid for randomly rough
surface with multiscale roughness, by comparing the theory predictions with the experimental results of
Sutter et al \cite{Sutter} for steel sliding on steel. The theory, which is based on the study of 
stress and temperature correlation functions, is valid for randomly rough surfaces with
roughness on arbitrary many decades in length scale. Within the uncertainty
of the experimental data (mainly the surface roughness power spectrum and the steel penetration hardness), 
there is good agreements between the theory and the experimental results.
\end{abstract}

\maketitle

\setcounter{page}{1}
\pagenumbering{arabic}

%\pagestyle{empty}

%%%%%%%%%%%%%% main text %%%%%%%%%%%%%%%%
%\begin{multicols}{2}

%%%%%%%%%%%%%% main text %%%%%%%%%%%%%%%%

{\bf 1 Introduction}

Friction between surfaces generates heat, leading to temperature increases at the contact points. 
This phenomenon is known as flash temperature, which is the high, localized, and brief temperature spike that 
occurs at the true points of contact between two rubbing solids. The rapid generation of heat at these contact 
points causes thermal spikes, resulting in intense flash temperatures that convert kinetic energy into 
thermal energy. These spikes can be extremely high, sometimes reaching 
over $1000^\circ {\rm C}$, but they are also incredibly brief,
lasting only for the instant that the asperities are in contact. 
The process is so rapid that the generated heat has little time to conduct away into the bulk of the materials, 
trapping thermal energy and further elevating the temperature at the contact point.

In almost all cases, most of the dissipated energy in sliding friction end up as thermal energy in the sliding 
solids. The temperature field in the solids can be written as $T({\bf x},t) = T_0({\bf x},t) + \Delta T({\bf x},t)$.
The {\it background} temperature $T_0({\bf x},t)$ varies slowly in space and time while the {\it flash} temperature
$\Delta T({\bf x},t)$ varies very fast in space and time. 
$\Delta T({\bf x},t)$ in non-zero only close to the asperity contact regions so very localized in space.

Frictional heating is important in very many applications, e.g., ice friction, rubber friction and
the friction between minerals. The flash temperature
can have a crucial influence on the friction force, usually reducing the friction force. This is the case for ice friction
and for granite sliding on granite, where the granite (mainly quartz) melting temperature may be reached 
at the sliding speeds (or order $\sim 1 \ {\rm m/s}$)
involved in earthquakes. Rubber friction depends (exponentially) on the temperature, 
and an increase in the temperature shift the friction coefficient master curve to higher sliding speeds, 
which usually reduce the friction but sometimes
increase it. For rubber friction relative advanced theories for the flash temperature was
developed in  Ref. \cite{rub1,rub2,rub3}.

In Ref. \cite{RiceHeat} Rice studied the importance of the flash heating on earthquake dynamics.
He considered a model where the frictional shear stress is constant until the flash temperature 
reaches a temperature of order the pseudotachylyte (rather than quartz or silica) melting temperature,
after which the shear stress was assumed to be negligible. However, the high stresses and temperatures
in the mineral contact regions are likely to to strongly weaken the interface during slip already
well before true melting occur\cite{rpp}. Minerals like quartz which are crystalline may during slip become amorphous
in the contact regions (quartz may locally converted to silica), and will soften continuously 
with increasing temperature or sliding speed \cite{rpp}.
This will result in a frictional shear stress which could decrease a lot even before true melting would occur,
as also observed for other crystalline materials, e.g., ice \cite{Norway}, which undergoes displacement-driven 
amorphization \cite{MarinIce}.

One way the flash temperature can manifest itself in experiments is  
as sliding-induced phase transformations in materials.
This was observed already in the classical studies by Bowden and Tabor, and more recently in many other studied \cite{mechano0,mechano1}.
However, in general these transformations may be mechanochemical in nature, involving {\it both} the high contact stresses and the flash temperature.
That is, the chemical or structural modifications result from stress aided thermal excitation's.

In this article I test a recently developed multiscale theory for the the flash temperature \cite{MP}.
The theory focus on temperature-stress correlation functions, which contain information about
the flash temperature. In the limiting case of roughness on one length scale the results are consistent with the classical
theories of Jaeger, Archard and Greenwood \cite{flash1,flash2,flash3} (see also \cite{flash4,flash5,flash6,TianGreen}).
%In Ref. \cite{TianGreen} Tian and Kennedy presented integral representations
%for the flash temperature focusing on square and circular moving heat sources.
However, for multiscale roughness covering several decades in length scale, the classical
theory fail severely\cite{MP}.

A multiscale flash temperature model was developed by Choudhry, Almqvist and Larsson \cite{flash6},
and tested in \cite{flashAlm2}, but this approach is purely numerical
can only be applied to surfaces with roughness over a short wavelength region. The approach developed in Ref. \cite{MP}
result in analytical equations which can be applied to systems with roughness over arbitrary many decades in wavelength.
Here the theory predictions are compared to experimental results for steel sliding on steel \cite{Sutter}.

In this study all temperatures refer to the {\it increase} in the temperature above the background temperature $T_0$.
Thus $T_{\rm flash}$ is a weighted average flash temperature,
and the actual temperature in the contact regions is $T_0+T_{\rm flash}$. Similarly,
temperature correlation functions like $\langle T({\bf x})T({\bf x}')\rangle$ are calculated with the background temperature
$T_0$ subtracted from $T({\bf x})$. Stated differently, all temperatures refer to actual temperatures if the background
temperature vanishes.

\begin{figure}
\includegraphics[width=0.3\textwidth,angle=0.0]{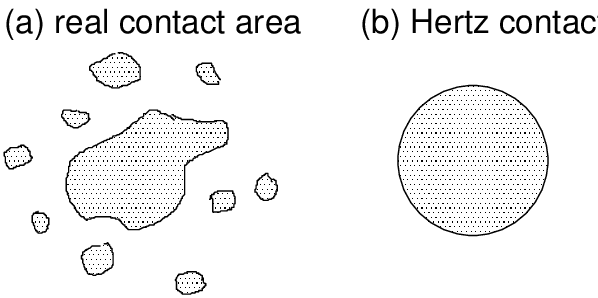}
\caption{\label{RealAndHertz.eps}
(a) A macroasperity contact area and (b) the contact are in Hertz approximation.
}
\end{figure}

\vskip 0.2cm
{\bf 2 Multiscale theory for the contact spot size and the flash temperature}

In Ref. \cite{Persson1,MP,Persson2,Persson3} we have derived analytical expressions for correlation functions involving the normal stress $\sigma ({\bf x})$ 
and the temperature $T({\bf x})$. These results can be used to estimate the size of the contact regions\cite{size,Persson2}, 
and the temperature distribution in the contact region. Here we summarize the most important results.

The stress-stress correlation function is defined by
$$g_\sigma (r) = \langle \sigma ({\bf x},0) \sigma({\bf 0},0)\rangle \eqno(1)$$
In Ref. \cite{Persson2,Persson3} we have shown that
$$g_\sigma (r) = (E^*)^2 {\pi \over 2} \int_{q_0}^{q_1} dq \, q^3 C(q) W(q) J_0(qr) \eqno(2)$$
where $E^* = E/(1-\nu^2)$ is the effective modulus, $J_0$ the 0-order Bessel function,
$C(q)$ is the surface roughness power spectrum, and
$$W(q) = P(q) [\gamma + (1-\gamma)P^2(q)]$$
where
$$P(q)={\rm erf}\left ({\sigma_0 \over 2 \surd G}\right )$$
$$G= {\pi \over 4} ( E^*)^2 \int_{q_0}^q dq \ q^3 C(q)$$
The surface has roughness components with wavenumbers $q_0 < q < q_1$ and the integrals over $q$ cover this wavenumber interval.

For a Hertzian-like contact with the diameter $2R$ the correlation function $g_\sigma(r)$ vanish for $r>2R$ \cite{Persson3}. 
For surfaces with multiscale roughness the macroasperity contact regions 
consist in general of a compact central part surrounded by smaller non-connected islands (see Fig. \ref{RealAndHertz.eps} and Ref. \cite{Mark,Persson3}),
which are elastically coupled, and for this case the $g(r)$ function will have a tail extending beyond the diameter of the central compact region.
For this case one can define an effective diameter of the macroasperities 
by the condition $g_\sigma (r)/g_\sigma (0) = \alpha$, where $\alpha <1 $ is a small positive number, e.g., $\alpha = 0.1$.

A multiscale theory for the flash temperature was developed in Ref. \cite{MP} and here we review the most important results.

In applications to sliding friction and wear it is the temperature in the asperity contact regions which matters,
not the temperature distribution outside of the contact regions. 
One can define an effective flash 
temperature in the asperity contact regions using
$$T_{\rm flash} = {\langle T({\bf x}) \sigma ({\bf x}) \rangle \over \langle \sigma ({\bf x}) \rangle}\eqno(3)$$
where $\langle \sigma ({\bf x}) \rangle = \sigma_0$ is the nominal (or average) stress.
Using the Persson contact mechanics theory one can show that \cite{MP}
$$ T_{\rm flash} = {\mu v \over \kappa \sigma_0} (E^*)^2 \int_{q_0}^{q_1} dq \, q^2 C(q) W(q)$$ 
$$\times {\rm Re} \int_0^{\pi/2} d\phi \, {1\over [1 -i (v/Dq) {\rm cos}\phi ]^{1/2}}\eqno(4)$$
where $\kappa$ is the thermal conductivity and $D=\kappa/ \rho C_{\rm p}$ the thermal diffusivity ($\rho $ is the mass density
and $C_{\rm p}$ the heat capacity).

The temperature distribution in the macroasperity contacts during sliding is anisotropic and larger
at the exit than at the front of the contact. This follows from the fact that the solid at the exit
is already heated-up by the contact with the asperity at the front of the contact. 
We can characterize this ``tilt'' of the temperature distribution using the correlation function
$$\nabla T_{\rm flash} ={\langle \nabla T({\bf x})  \sigma ({\bf x})\rangle \over \langle \sigma ({\bf x}) \rangle} .\eqno(5)$$
Choosing the $x$-axis along the sliding direction only the $x$-component of will be non-vanishing which we write 
$$T'_{\rm flash} = {\langle \partial_x T({\bf x})  \sigma ({\bf x})\rangle \over \langle \sigma ({\bf x}) \rangle }, \eqno(6)$$
where $\partial_x f = \partial f/ \partial x$, and
where the $'$ indicate derivative of $T({\bf x})$ with respect to $x$. We get
$$ T'_{\rm flash} = {\mu v \over \kappa \sigma_0} (E^*)^2 \int_{q_0}^{q_1} dq \, q^3 C(q) W(q) $$
$$\times {\rm Re} \int_0^{\pi/2} d\phi \, {i {\rm cos}\phi \over [1 -i (v/Dq) {\rm cos}\phi ]^{1/2}} \eqno(7)$$
Since the flash temperature is higher at the exit than at the leading edge of the moving contact region,
the temperature profile tilt upwards towards the exit of the contact regions. If the positive direction of the $x$-axis is along the sliding direction
then $T'_{\rm flash} <0$ for a moving
contact region, while for a stationary contact by symmetry $T'_{\rm flash} = 0$.

It is easy to calculate similar averages of higher order derivatives of $T({\bf x})$. The most important
quantity is $ \nabla^2 T_{\rm flash}$, which can be used to estimate the size of the flash temperature hot spots. 
We define
$$ \nabla^2 T_{\rm flash} = {\langle \nabla^2 T ({\bf x}) \sigma ({\bf x}) \rangle \over \langle \sigma ({\bf x}) \rangle }\eqno(8)$$ 
we get
$$ \nabla^2 T_{\rm flash} = -{\mu v \over \kappa \sigma_0} (E^*)^2 \int_{q_0}^{q_1} dq \, q^4 C(q) W(q) $$
$$\times {\rm Re} \int_0^{\pi/2} d\phi \, {1\over [1 -i (v/Dq) {\rm cos}\phi ]^{1/2}}\eqno(9)$$

\begin{figure}
\includegraphics[width=0.3\textwidth,angle=0.0]{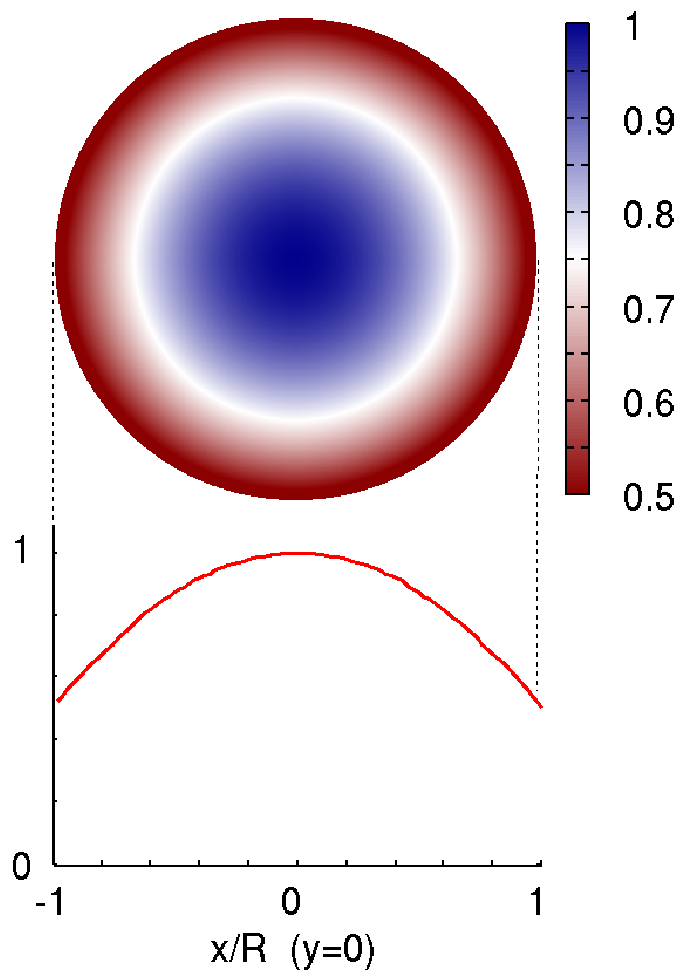}
\caption{\label{picINTENSITYstationary.eps}
The temperature distribution in a stationary Hertz contact.
The top is a $x,y$ 2D view and the bottom curve the temperature profile through
the center line $y=0$. 
}
\end{figure}

\vskip 0.2cm
{\bf 3 Sliding Hertz contact}

When observed at low magnification a macroasperity may appear smooth, and the contact region
Hertzian-like, and it is therefore interesting to 
correlate the results of the full theory in Sec. 2 with the known results for the
sliding Hertz contact. For this case we will also calculate $\langle \nabla^2 T \rangle$
and $\langle \partial_x T \rangle$ which have not been obtained before.
We will only consider the case of stationary contact, and sliding contact for high sliding
speed.

If we approximate a macroasperity contact region as a
Hertz contact with the pressure distribution
$$\sigma (r) = \sigma_1 \left [1-\left ( {r \over R}\right )^2 \right ]^{1/2} \eqno(10)$$ 
then the heat source 
$$\dot q (r) = \dot q_1 \left [1-\left ( {r \over R}\right )^2 \right ]^{1/2} \eqno(11)$$
where
$$\dot q_1  = {3 \dot Q \over 2 \pi R^2} \eqno(12)$$
where $\dot Q$ is the total power produced by the heat source. For a stationary contact
the heat diffusion equation can be easily solved. The temperature field for $r<R$ is given by
$$T (r) = T_1 \left [1- {1\over 2} \left ( {r \over R}\right )^2 \right ]\eqno(13)$$
where  $T_1 = \dot q_1 R \pi/4\kappa $. Fig. \ref{picINTENSITYstationary.eps} shows  $T(r)/T_1$ 
for this temperature distribution.

In what follows $\langle .. \rangle$ denote integrating over the surface area.
Using the results above we obtain the weighted average temperature:
$$T_{\rm flash} = {\langle T ({\bf x}) \sigma ({\bf x}) \rangle \over \langle \sigma ({\bf x}) \rangle } 
= T_1 {\int_0^1 dr \, r \left (1-r^2 \right )^{1/2} \left (1- r^2 /2 \right ) \over \int_0^1 dr \, r \left (1-r^2 \right )^{1/2} } $$
$$ =0.8 T_1 = 0.3 {\dot Q \over \kappa R} \eqno(14)$$
The maximum temperature is
$$T_{\rm max} = T_1 = {3\over 8} {\dot Q \over \kappa R} \approx 0.375 {\dot Q \over \kappa R}\eqno(15)$$
and the average is
$$T_{\rm av} =  {1\over \pi R^2} \int d^2x \, T({\bf x}) = {3 \over 4} T_1 \approx 0.281 {\dot Q \over \kappa R} \eqno(16)$$
The flash temperature is, as expected, between $T_{\rm av}$ and $T_{\rm max}$.

Next, since
$$\nabla^2 T  = - {2 \over R^2} T_1 $$
we get
$$\nabla^2 T_{\rm flash} = {\langle \nabla^2 T ({\bf x}) \sigma ({\bf x}) \rangle  \over \langle \sigma({\bf x}) \rangle } = - {2 \over R^2} T_1 \eqno(17)$$
and
$${\nabla^2 T_{\rm flash} \over T_{\rm flash}} = {\langle \nabla^2 T ({\bf x}) \sigma ({\bf x})\rangle  \over \langle T ({\bf x})\sigma ({\bf x}) \rangle} =- {2.5 \over R^2}\eqno(18)$$
Finally, since the temperature distribution is independent of the angular coordinate $\phi$, the inverse of the 
slope length vanish,
$${1\over d_{\rm flash}} = {T'_{\rm flash}\over T_{\rm flash}} = 0$$

\begin{figure}
\includegraphics[width=0.3\textwidth,angle=0.0]{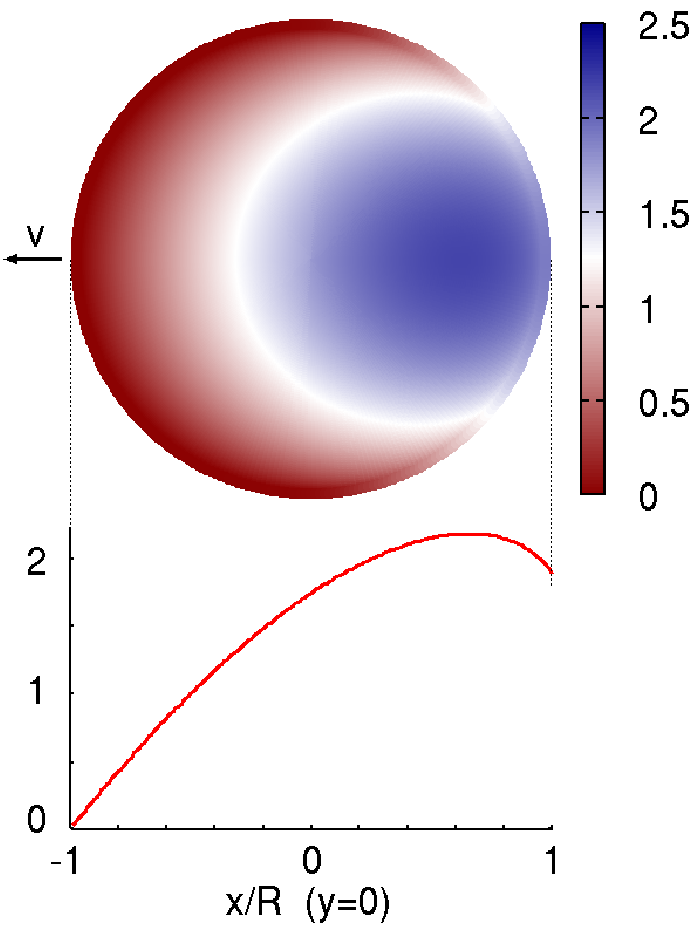}
\caption{\label{1x.2Intensity.eps}
The temperature distribution in a fast moving Hertz contact.
The top is a $x,y$ 2D view and the bottom curve the temperature profile through
the center line $y=0$. 
}
\end{figure}

Next, let us consider high sliding speeds. In this case one can neglect the lateral
diffusion of the heat. Measuring the coordinates $x$ and $y$ in units of the radius $R$ 
the temperature is given by the integral representation (see Appendix A):
$$T(x,y) = {\dot q_1 \over \kappa} \left ({D R \over \pi v} \right )^{1/2} f(x,y) \eqno(19)$$
where
$$f(x,y)= \int_{-s}^x d\xi \, { (1-\xi^2 -y^2)^{1/2}  \over (x-\xi)^{1/2}} \eqno(20)$$
where $s=(1-y^2)^{1/2}$. An intensity map of $f(x,y)$ is shown in Fig. \ref{1x.2Intensity.eps}.

Using (19) and (20) one can calculate all quantities of interest using numerical integration.
Here we summarize the most important results (see also Appendix A).

The flash temperature is given by
$$T_{\rm flash} \approx 0.352 {\dot Q \over R \kappa} \left ({D  \over R v} \right )^{1/2} \eqno(21)$$
The maximum temperature is
$$T_{\rm max} \approx 0.590 {\dot Q \over R \kappa} \left ({D  \over R v} \right )^{1/2} \eqno(22)$$
and the average flash temperature is 
$$T_{\rm av} \approx 0.323 {\dot Q \over R \kappa} \left ({D  \over R v} \right )^{1/2} \eqno(23)$$
which is a factor of $\approx 0.549$ smaller than the maximal temperature. The flash temperature
is, as expected, between $T_{\rm av}$ and $T_{\rm max}$.

In the Appendix A we show that
$${\nabla^2 T_{\rm flash} \over T_{\rm flash}} = {\langle \nabla^2 T ({\bf x}) \sigma ({\bf x})\rangle  
\over \langle T ({\bf x}) \sigma ({\bf x}) \rangle} \approx -{4.93 \over R^2} \eqno(24)$$
Note that the prefactor $4.93$ is larger than in the static case where it equals 2.5. This is intuitively plausible from the intensity
maps in Fig. \ref{picINTENSITYstationary.eps} and \ref{1x.2Intensity.eps}, 
where the hottest region is more concentrated in space for high sliding speeds. 

We define the width $D_{\rm flash}$ and the slope-length $d_{\rm flash}$ by
$$D_{\rm flash} = \left [ {-T_{\rm flash} \over \nabla^2 T_{\rm flash}} \right ]^{1/2}\eqno(25)$$
$$d_{\rm flash} = {-T_{\rm flash} \over T'_{\rm flash}}\eqno(26)$$
For sliding Hertz contact: 
$$D_{\rm flash} = \alpha R$$
where, using (18), $\alpha \approx 0.63$ for very low velocity (stationary contacts) and, using (24), $\alpha \approx 0.45$ for very high velocities.
Here high and low velocities refer to $v \gg v^*$ and $v \ll v^*$, with $v^* = D/R$, respectively.

In the Appendix A we derive the slope parameter for the sliding Hertzian contact:
$$d_{\rm flash} \approx 1.36 R$$
while $d_{\rm flash}=\infty$ for $v=0$. Note that for high sliding speed
$d_{\rm flash} /D_{\rm flash} \approx 1.36/0.45 \approx 3.0$.

\begin{figure}
\includegraphics[width=0.3\textwidth,angle=0.0]{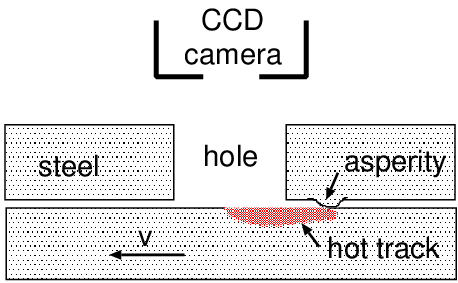}
\caption{\label{HoleHotTrack.eps}
The hot track on the sliding steel surface from the contact with an asperity
on the stationary surface located close to the hole (schematic).
}
\end{figure}

\begin{figure}
\includegraphics[width=0.47\textwidth,angle=0.0]{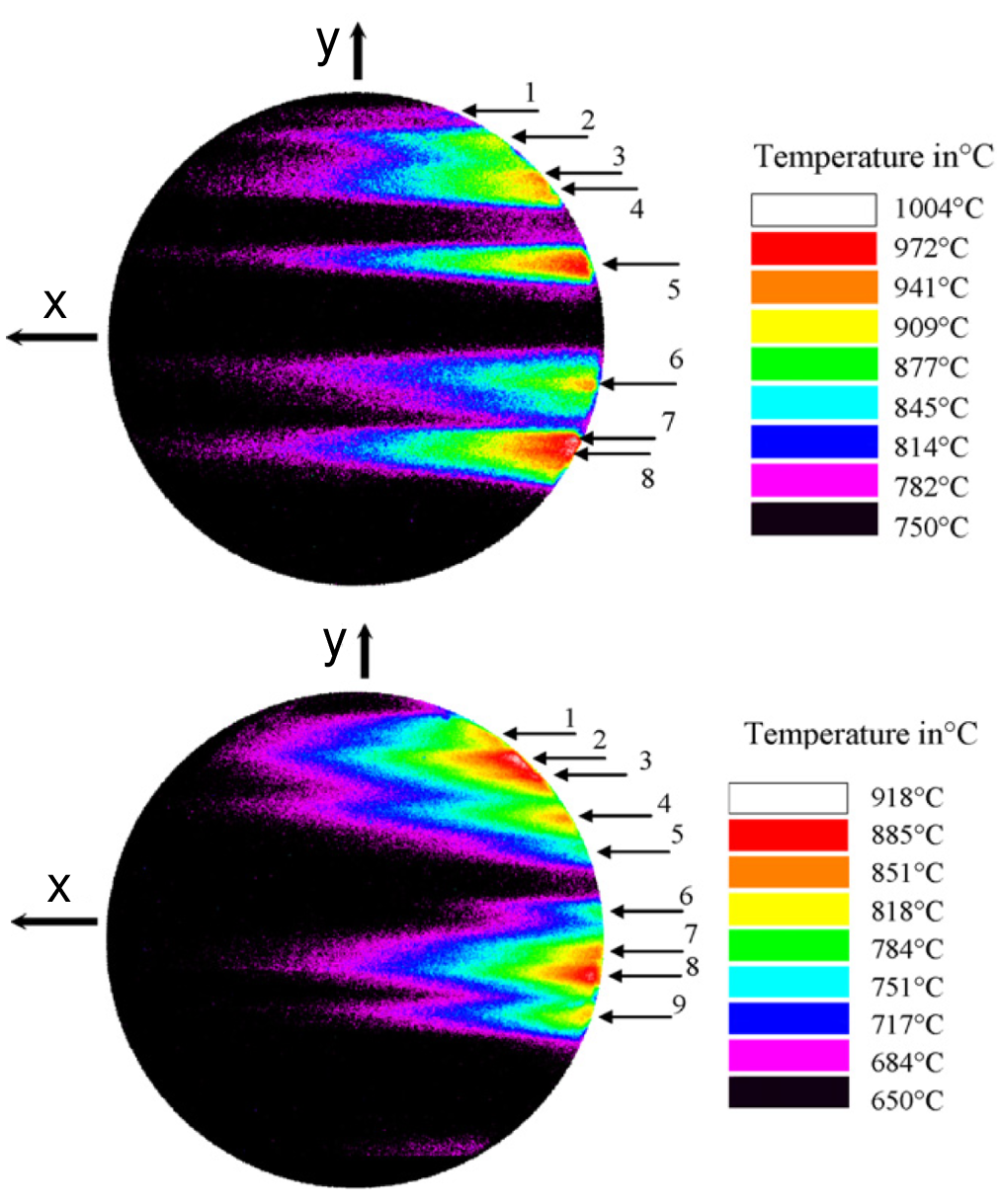}
\caption{\label{TempInHole.eps}
Thermography of the sliding surface at the sliding speed (a) $33.9 \, {\rm m/s}$ (test 3) and
(b) $34.5 \, {\rm m/s}$ (test 5). 
Adapted from Ref. \cite{Sutter}.
}
\end{figure}

\begin{figure}
\includegraphics[width=0.47\textwidth,angle=0.0]{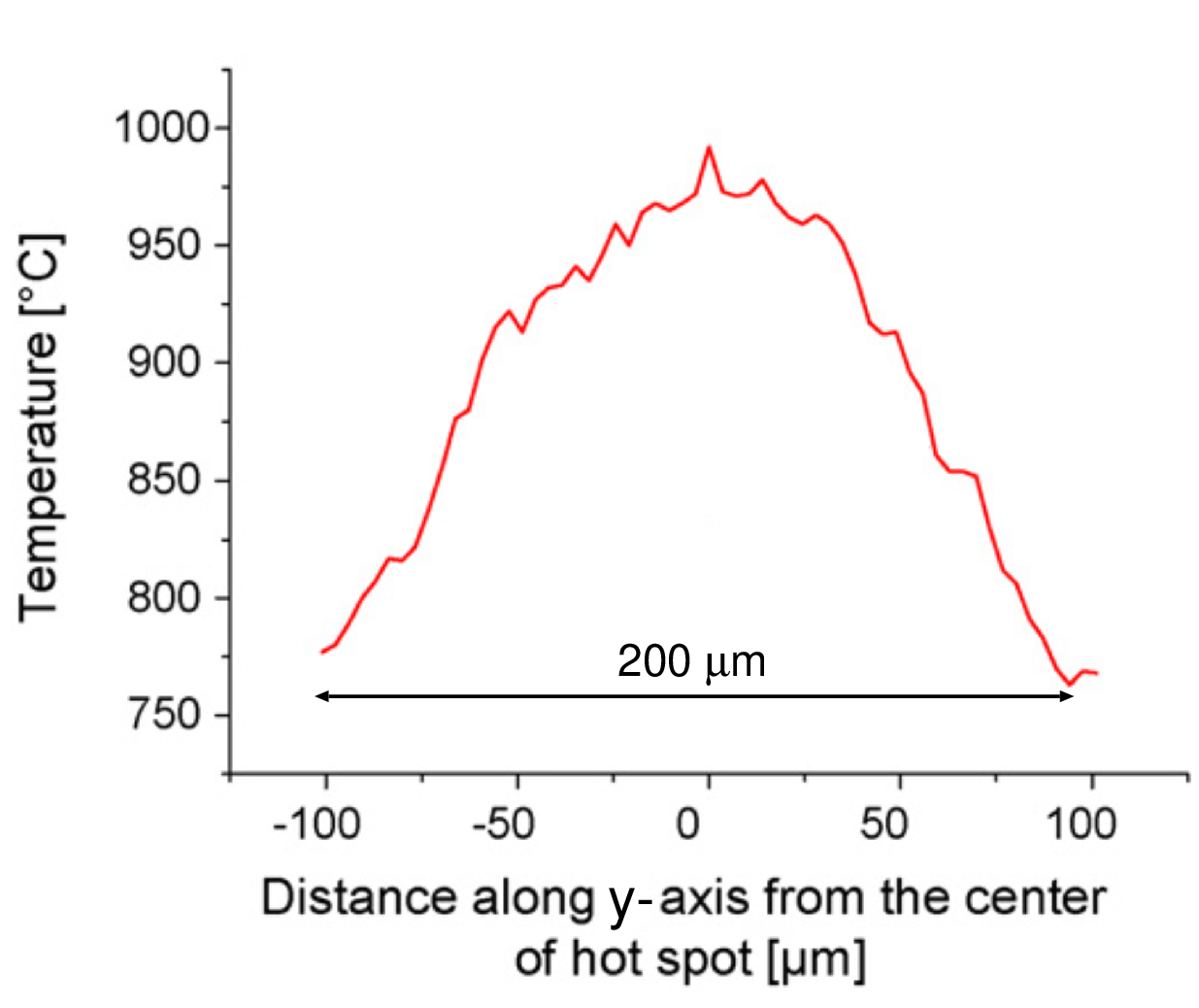}
\caption{\label{HotSpotProfileSteel.eps}
Spatial temperature distribution along $x$-axis near point 5 for test 3.
Adapted from Ref. \cite{Sutter}.
}
\end{figure}

%\begin{figure}
%\includegraphics[width=0.47\textwidth,angle=0.0]{TempAlongDistanceAway.eps}
%\caption{\label{TempAlongDistanceAway.eps}
%Temperature distribution along $y$-axis for test 3 at point 5, 7 and 2.
%}
%\end{figure}

\vskip 0.2cm
{\bf 4 Numerical results and comparison with experiments}

The flash temperature was studied experimentally for steel sliding on steel
and here we will compare the theory predictions (Sec. 2) with the experimental data presented in Ref. \cite{Sutter}.
We first briefly review the results obtained in Ref. \cite{Sutter}. The experiments consisted of a steel
block (substrate) sliding against against a stationary steel block with a small hole (radius $1 \ {\rm mm})$ drilled in it.
During sliding the asperities on the stationary block will result in hot tracks on the substrate surface which can be observed
through the hole using a CCD camera as illustrated schematically in Fig. \ref{HoleHotTrack.eps}. 
After calibration of the camera the detailed surface temperature
profile can be measured with the spatial resolution $4 \ {\rm \mu m}$. Because of thermal diffusion, 
only the hot tracks from asperities on the stationary block, which contact the substrate close to the hole, 
can be observed, and the most hot tracks will result from asperity contacts very close to the hole.
Thus the width of hottest tracks at the edge of the hole will give an estimation of the effective width of the contact
patches, which we will compare to the theory presented above.

Fig. \ref{TempInHole.eps} shows the thermography of the hot tracks in two cases
where the sliding speed was $33.9 \, {\rm m/s}$ (test 3) and $34.5 \, {\rm m/s}$ (test 5).
Fig. \ref{HotSpotProfileSteel.eps} shows the spatial temperature distribution 
along $y$-axis (orthogonal to the sliding direction) near point 5 for test 3. For this test
the asperity contact is assumed to be just outside the hole so this figure gives an estimation of the
width of the hot spots. Similar width was obtained for other asperity contacts when the asperities
was located very close to the boundary of the drilled hole. 

\begin{figure}
\includegraphics[width=0.47\textwidth,angle=0.0]{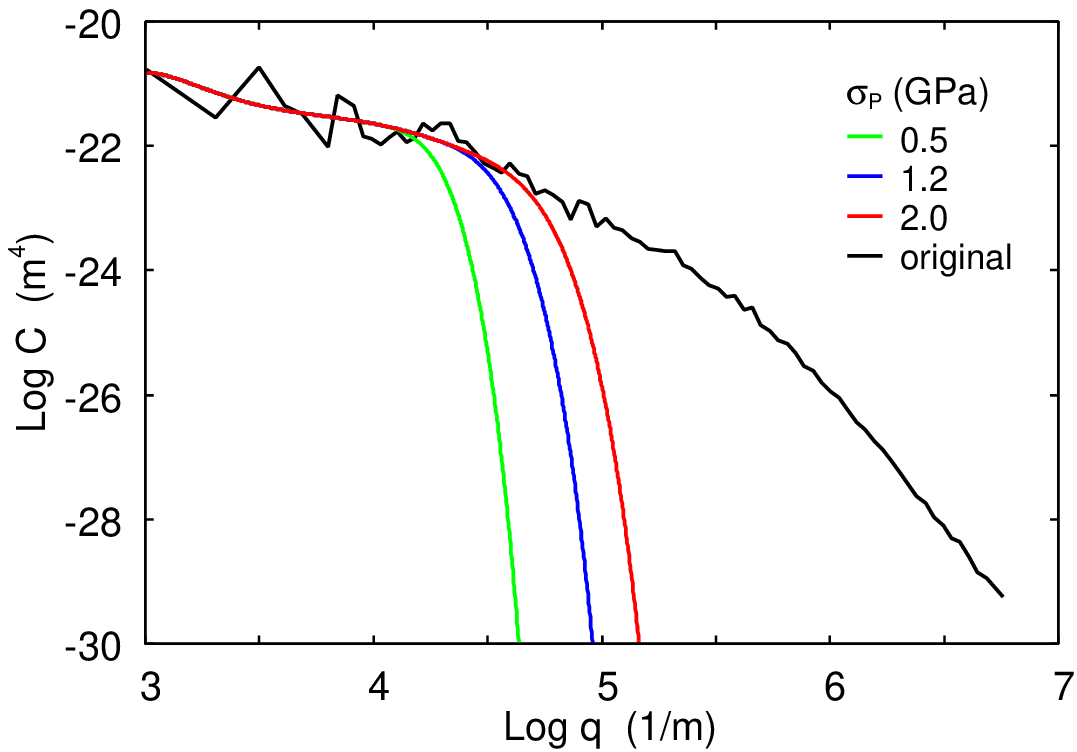}
\caption{\label{1logv.2logC.steel.yielded.eps}
The black line is the surface roughness power spectrum of a machined steel surface scaled by a factor of 2.
The original (not scaled) surface has the
root-mean-square roughness $h_{\rm rms} \approx 0.90 \ {\rm \mu m}$ and the rms slope $\xi \approx 0.30$.
The green, blue and red lines are the power spectrum of the plastically smoothed profiles using the penetration hardness
$\sigma_{\rm P} = 0.5$, 1.2 and $2 \, {\rm GPa}$, respectively.
}
\end{figure}

We will analyze the experimental data using the theory described in Sec. 2.
The power spectrum of the surface roughness of the steel surfaces was not given in Ref. \cite{Sutter}
but only the ${\rm R_a}$ roughness parameter was given, ${\rm R_a} \approx 0.8 \ {\rm \mu m}$. In the study below
we use the surface roughness power spectrum $C(q)$ of a machined steel surface with 
similar roughness as in Ref. \cite{Ruibin}. The black line in Fig. \ref{1logv.2logC.steel.yielded.eps} gives the power spectrum
of this steel surface.

For the medium-low carbon steel C22 used in Ref. \cite{Sutter},
we use the following elastoplastic parameters: Young's modulus $E=200 {\rm GPa}$, Poisson ratio $\nu = 0.3$
and for the penetration hardness three values, namely $\sigma_{\rm P} = 0.5$, $1.2$ and $2.0 \ {\rm GPa}$. The Brinell hardness
quoted in Ref. \cite{Sutter} was 120 (corresponding to $\sigma_{\rm P} \approx 1.2 \ {\rm GPa}$) but the stress 
needed for plastic deformation of asperities may differ from the macroscopic penetration hardness for reasons discussed
elsewhere\cite{Metal1}. Here we therefore use three values for $\sigma_{\rm P}$ to show the dependency (sensitivity) 
of the results to the penetration hardness.
To take into account the contact between two steel surfaces we use the effective modulus $E/2 = 100 \ {\rm GPa}$ 

We use the Persson contact mechanics theory\cite{Persson}. We assume that the material deform plastically without strain hardening in such a way that the maximum
stress in the contact areas is $\sigma_{\rm P}$. For this elastoplastic model
it has been shown that the theory accurately predict both the elastic and plastic 
contact area \cite{plast1}. Since the theory discussed in Sec. 2 assumes elastic deformation we replace the original surface with a surface
where the short wavelength roughness is smoothed by the plastic deformations.
This smoothing is done everywhere, i.e. also in the non-contact area, which is necessary in order to have a randomly rough surface also after the smoothing.

Plastic smoothing can be done in different ways and here we use a simple procedure which gives a smoothed surface
where the (elastic) contact area as a function of magnification is virtually the same 
as the elastoplastic contact area using the original surface\cite{explain}. 
The basic idea is that if the applied stress is removed and then applied again the surface will deform elastically as long as the applied
stress is smaller than the stress used to plastically deform the surface.

The power spectrum $C_{\rm sm} (q)$ of the smoothed surface is obtained from the power spectrum
$C(q)$ of the original surface using:
$$C_{\rm sm} (q) = \left [1-\left ({A_{\rm pl} (\zeta) \over A_{\rm pl0}} \right )^6 \right ] C(q) .\eqno(27)$$
Here $A_{\rm pl} (\zeta)$ is the plastic surface area predicted by the elastoplastic calculation when the interface
is observed at the magnification $\zeta = q/q_0$ (where $q_0$ is the smallest wavenumber) and $A_{\rm pl0}= F_{\rm N}/\sigma_{\rm P}$
the plastic contact area observed at the highest magnification where all contact regions have yielded plastically.
The smoothing (27) consist of removing or reducing the amplitude of the short wavelength roughness components which deform plastically. 

The black line in Fig. \ref{1logv.2logC.steel.yielded.eps} shows the surface roughness power spectrum of the original steel surface 
scaled by  a factor of 2 to take into account the contact between two rough surfaces with identical power spectra but uncorrelated roughness.
The green, blue and red lines are the power spectrum of the plastically smoothed profiles using the penetration hardness $\sigma_{\rm P} = 0.5$,
$1.2$ and $2 \, {\rm GPa}$, respectively.

\begin{figure}
\includegraphics[width=0.47\textwidth,angle=0.0]{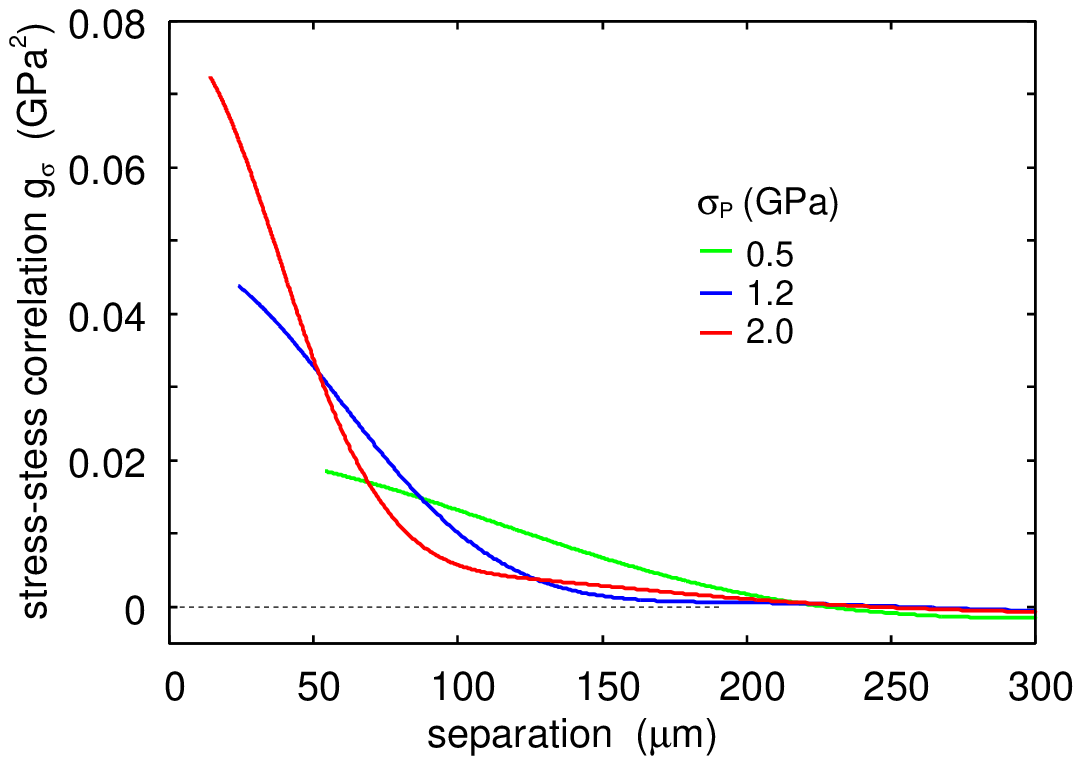}
\caption{\label{1distance.2stessStress.steel.eps}
The stress-stress correlation function $g_{\rm sigma} (r) = \langle \sigma ({\bf x}) \sigma ({\bf x}') \rangle$
as a function of the separation $r=|{\bf x}-{\bf x}'|$ between the points ${\bf x}$ and ${\bf x}'$ for the surface plastically smoothed
assuming the penetration hardness $\sigma_{\rm P} = 0.5$ (green line), $1.2$ (blue line) and $2 \, {\rm GPa}$ (red line).
}
\end{figure}

Fig. \ref{1distance.2stessStress.steel.eps} shows the
stress-stress correlation function $g_{\rm \sigma} (r) = \langle \sigma ({\bf x}) \sigma ({\bf x}') \rangle$
as a function of the separation $r=|{\bf x}-{\bf x}'|$ between the points ${\bf x}$ and ${\bf x}'$ for the surface plastically smoothed
assuming the penetration hardness $\sigma_{\rm P} = 0.5$ (green line), $1.2$ (blue line) and $2 \, {\rm GPa}$ (red line).
For a Hertzian contact $g_{\rm \sigma} (r) = 0$ for separations $r$ larger than the diameter $2R$ of the contact area \cite{Persson3}. In the present case
the contacts are not circular, and island-like contact regions will surround the main compact inner region (see Fig. \ref{RealAndHertz.eps})
and for this case there is not a sharp cut-off in the $g_{\rm \sigma} (r)$ function. Nevertheless, we can define an effective radius of the macroasperity
contact area using $g_{\rm \sigma} (r)/ g_{\rm \sigma} (0) \approx 0$. Using such a definition the figure shows that the
diameter of the macroasperity contact regions is of order $\approx 200 \ {\rm \mu m}$ if $\sigma_{\rm P}$ is chosen as the Brinell hardness.

\begin{figure}
\includegraphics[width=0.47\textwidth,angle=0.0]{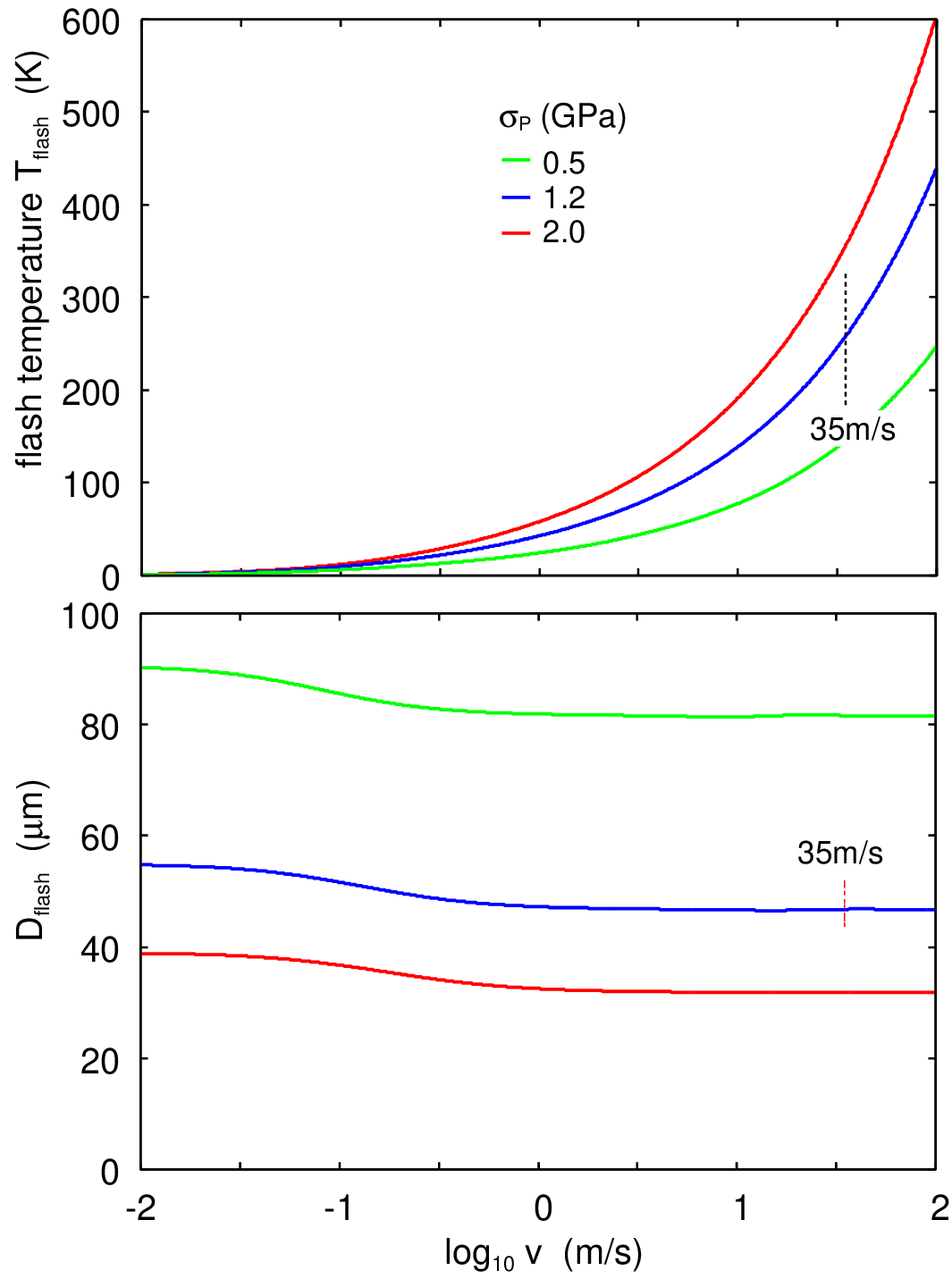}
\caption{\label{1logv.2Tflash.steel.eps}
(a) The weighted average flash temperature $T_{\rm flash}$, and (b) the effective diameter $D_{\rm flash}$ of the hot spots,
as a function of the logarithm of the sliding speed for the three plastically smoothed surfaces.
}
\end{figure}

Next we focus on the result of the calculation of the flash temperature.
In what follows we use the measured nominal contact pressure $\sigma_0 \approx 90 \, {\rm MPa}$,
and the following thermal parameters for steel:
mass density $\rho = 7900 \, {\rm kg/m^3}$, thermal conductivity $\kappa = 30.0 \, {\rm W/Km}$
and the thermal heat capacity $C_{\rm P} = 500.0 \, {\rm J/kgK}$. We note that the thermal conductivity at room temperature for the used
steel is about $50.0 \, {\rm W/Km}$, but for the higher temperatures relevant here it is about $30  \, {\rm W/Km}$.
(The thermal conductivity of materials decreases at high temperatures due to increased scattering 
of the thermal carriers. Thus, as temperature rises, 
the phonon and electron mean free path becomes shorter, which reduces the efficiency of heat transfer.) 

Here we note that Reddyhoff et. al. \cite{ReddyhoffHardening} have shown that
the thermal conductivity of solids (mainly metals) close to the surface
may differ from that in the bulk. Thus work hardening, which produce disordered lattice structures 
and defects, e.g., dislocations or point defects, decreases the surface thermal conductivity.
This is due to scattering of the thermal carriers (phonons and electrons) from the lattice imperfections,
which result in a reduced carrier mean free path and smaller thermal conductivity.
These defects, however, have only a small influence on the
mass density and the thermal heat capacity, and these quantities are expected to be nearly the same in the surface region as in the bulk.

Reddyhoff et al found that the measured thermal conductivity of AISI 52100 steel
($21 \ {\rm W/mK}$) is less than half the value cited in the literature ($46 \ {\rm W/mK}$). 
This discrepancy arises from a reduction in thermal conductivity of AISI 52100 due to work-hardening. They pointed out that the
thermal conductivity value generally cited and used in the literature represents that of soft, annealed alloys, but work
hardened AISI 52100, which is generally employed in rolling bearings and for lubricant testing, appears to have a much
lower surface thermal conductivity.

The measured friction coefficient was not constant during the sliding tests and here we use an average $\mu \approx 0.11$. This friction coefficient
is much smaller than typically found for steel sliding on steel at low sliding speeds 
(typically $\mu \approx 0.7$ for $v \approx 1 \ {\rm mm/s}$ \cite{NATURE}),
and is likely a result of the high temperatures $\sim 1000^\circ {\rm C}$ prevailing in the
asperity contact regions, which result from the high sliding speed (about $35 \ {\rm m/s}$).
If one assume that $T_{\rm flash}$ is continuous at the interface then if $s$ is the fraction of energy going into the sliding block (substrate)
and $1-s$ into the stationary block then $(1-s)T_{\rm flash} ({\rm stationary}) = s T_{\rm flash} ({\rm moving})$, where
$T_{\rm flash} ({\rm stationary})$ and $T_{\rm flash} ({\rm moving})$ are calculated 
assuming that {\it equal} thermal energy goes into each block. Using this equation 
for the sliding speed used in the experiments, $v=35 \ {\rm m/s}$, assuming the penetration hardness $\sigma_{\rm P} = 1.2 \ {\rm GPa}$,
the theory predict that the fraction of the frictional energy going into the sliding block is 
$s\approx 0.938$ and the rest into the stationary block. In what follows we reduce the frictional energy by a factor of $s\approx 0.938$ in order
to take this effect into accoint when calculating the flash temperature on the sliding substrate surface.

\begin{figure}
\includegraphics[width=0.47\textwidth,angle=0.0]{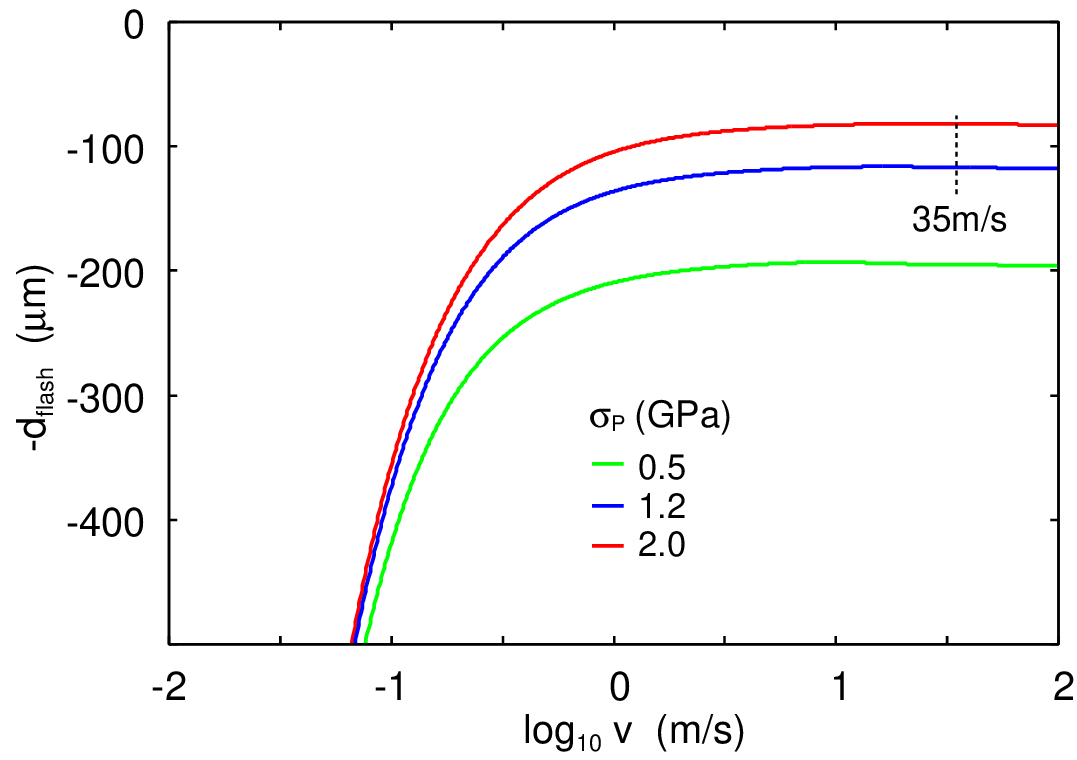}
\caption{\label{1logv.2tiltLENGTH.steel.eps}
The temperature slope-length $d_{\rm flash}$
as a function of the logarithm of the sliding speed for the three plastically smoothed surfaces.
}
\end{figure}

Fig. \ref{1logv.2Tflash.steel.eps}(a) shows the calculated flash temperature $T_{\rm flash}$, and (b) the width $D_{\rm flash}$,
as a function of the logarithm of the sliding speed for the three plastically smoothed surfaces. For the sliding speeds in the experiments (about $35 \ {\rm m/s}$),
indicated by the vertical dotted lines, the temperature increase is $259^\circ {\rm C}$ for the yield stress $1.2 \ {\rm GPa}$. The measured temperature increase is
about $\sim 200 \ {\rm K}$, but the actual maximum temperature increase may be slightly larger since the temperature 
measurements are for some short distance away from the hot spots.
The width parameter $D_{\rm flash} = 46.58 \ {\rm \mu m}$ for the yield stress $1.2 \ {\rm GPa}$. 
Assuming the result (25) derived for a sliding Hertz contact with $\alpha = 0.45$ (since $v >> v^*$ in the present case) gives $2R \approx 207 \ {\rm \mu m}$.
This is similar to the diameter as predicted from the stress correlation function and
also observed in the experiments. Finally, in Fig. \ref{1logv.2tiltLENGTH.steel.eps} we show the slope-length $d_{\rm flash}$ 
as a function of the logarithm of the sliding speed for the three plastically smoothed surfaces. For the sliding speed $35 \ {\rm m/s}$ this gives
$118 \ {\rm \mu m}$ if $\sigma_{\rm P} = 1.2 \ {\rm GPa}$. For Hertzian contact the theory
predict the slope length $d_{\rm flash} \approx 0.68 \times [2R]$ so that if the diameter is $\approx 200 \ {\rm \mu m}$ then the Hertz contact theory
would predict $136 \ {\rm \mu m}$ which is consistent with the full theory. Hence it appears that the sliding Hertz contact results gives useful results for the present system.
The reason for this is that plastic deformation occur already at relative long length scales, and only roughness over a relative narrow
length-scale region matters, which is the limit where the classical flash temperature descriptions are approximately valid (see Ref. \cite{MP}).
However, when roughness on many length scales are important the classical theory fail even qualitatively, as was 
the case for granite sliding on granite in Ref. \cite{MP}.

Finally, we note that Fig. \ref{TempInHole.eps} shows that the surface temperature away from the hot tracks behind the asperity contact regions
is about $700^\circ {\rm C}$ i.e., much higher that room temperature. This background temperature is the cumulative result of 
the flash temperature. It formally correspond to the $q=0$ contribution in the Fourier decomposition of the temperature profile.
It can be calculated assuming a uniform heat source $\dot q_0 = \mu \sigma_0 v$ with an onset at the time $t=0$ of start of sliding. This result in
a background surface temperature given by
$$T_0 (t) = {1\over \kappa} \left ({D\over \pi} \right )^{1/2} \int_0^t dt' \, {\dot q (t') \over (t-t')^{1/2}} = {2 \dot q_0 \over \kappa} 
\left ({D\over \pi} \right )^{1/2} \surd t $$
In reality, the background temperature is also influenced by heat transfer to the surrounding gas but this may not be so important
in the present case due to the short sliding time.

\vskip 0.2cm
{\bf 5 Summary and conclusion}

In this study I have tested a recently developed theory for the 
size of the contact regions, and for the flash temperature, for solid blocks in sliding contact \cite{MP}.
The theoretical predictions are consistent with the experimental observations by
Sutter et al \cite{Sutter} for steel sliding on steel. However, an ``exact'' comparison between theory and experiments
is not possible because the surface roughness power spectrum of the steel surfaces used in Ref. \cite{Sutter} is not
known, and also the (macroscopic) Brinell harness may differ somewhat from the penetration hardness
involved in plastic deformations of asperities at the micrometer length scale \cite{Metal1}.
The theory is based on multiscale contact mechanics, and is valid for surfaces
with roughness on an arbitrary number of decades in length scale. 

We have found that the flash temperature parameters $D_{\rm flash}$ and $d_{\rm flash}$ 
for the sliding Hertz heat source are consistent with the the result of the full theory 
and with the experiments for steel sliding on steel \cite{Sutter}. 
The reason for this is that the steel deform plastically already at relative long length scales
so that effectively roughness occur only on a relative narrow wavelength region, and in particular
the short wavelength roughness on the original surfaces is irrelevant for the flash temperature.
However, this is not the case in general, and in Ref. \cite{MP} it was shown that the
classical theory for the flash temperature fail severely for surfaces with roughness 
extending over 2 decades or more of length scale. This was the case for granite sliding 
on granite \cite{MP}, a system of interest in earthquake dynamics.

\vskip 0.2cm
{\bf Appendix A: Sliding Hertzian contact}

\begin{figure}
\includegraphics[width=0.3\textwidth,angle=0.0]{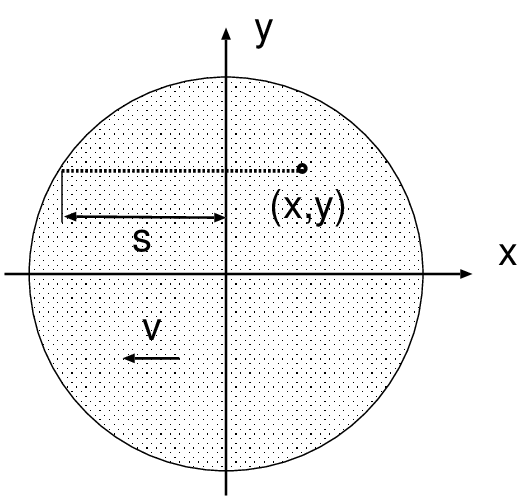}
\caption{\label{FlashCircle.eps}
The Hertz circular contact region moves with the velocity $v$ 
to the left along the negative $x$-axis. A point $(x,y)$ has been in contact
with the heat source for the time $t=(s+x)/v$ or the sliding distance $s+x$ (dotted line).
}
\end{figure}

We consider a sliding Hertz contact in the limit of very
high sliding speeds where one can neglect the lateral
diffusion of the heat. In this case, if a material point experience the time dependent heat source
$\dot q (t)$ for $t>0$, then the temperature becomes
$$T(t) = {1\over \kappa} \left ({D\over \pi} \right )^{1/2} \int_0^t dt' \, {\dot q(t') \over (t-t')^{1/2}  }$$
Fig. \ref{FlashCircle.eps} shows the Hertz circular contact region moving with the velocity $v$ 
along the negative $x$-axis. A point $(x,y)$ has been in contact
with the heat source for the time $t=(s+x)/v$ or the distance $s+x$ (dotted line).
If we write $x' = -s+vt'$, where $s= (R^2-y^2)^{1/2}$, we get
$$T(t) = {1\over \kappa} \left ({D\over \pi v} \right )^{1/2} \int_{-s}^x dx' \, {\dot q_1  \over (x-x')^{1/2}} \left (1- {x'^2+y^2 \over R^2} \right )^{1/2}$$
Writing $x'=R \xi$ and measuring $x$ and $y$ in units of $R$ gives
$$T(x,y) = {\dot q_1 \over \kappa} \left ({D R \over \pi v} \right )^{1/2} f(x,y)\eqno(A1)$$
where
$$f(x,y)= \int_{-s}^x d\xi \, { (1-\xi^2 -y^2)^{1/2}  \over (x-\xi)^{1/2}}\eqno(A2)$$
where $s=(1-y^2)^{1/2}$. An intensity map of $f(x,y)$ is shown in Fig. \ref{1x.2Intensity.eps}. 
Using (A1) and (A2) one can calculate
$$T_{\rm flash} =  {\langle T ({\bf x}) \sigma ({\bf x}) \rangle \over \langle \sigma ({\bf x}) \rangle } 
= 1.305 {\dot q \over \kappa} \left ({D R\over \pi v} \right )^{1/2}$$ 

Using $\dot q_1 = 3\dot Q / 2 \pi R^2$ we can write
$$T(x,t) = {\dot Q \over R \kappa} \left ({D  \over R v} \right )^{1/2} {3 \over 2 \pi \surd \pi} f(x,y)$$
and
$$T_{\rm flash} \approx 0.352 {\dot Q \over R \kappa} \left ({D  \over R v} \right )^{1/2} \eqno(A3)$$
The maximum temperature is
$$T_{\rm max} \approx 0.590 {\dot Q \over R \kappa} \left ({D  \over R v} \right )^{1/2} \eqno(A4)$$
and the average flash temperature is 
$$T_{\rm av} \approx 0.323 {\dot Q \over R \kappa} \left ({D  \over R v} \right )^{1/2} \eqno(A5)$$
which is a factor of $\approx 0.549$ smaller than the maximal temperature. The flash temperature
is, as expected, between $T_{\rm av}$ and $T_{\rm max}$.

Next we calculate $\nabla^2 T_{\rm flash}$. We write
$$\langle \nabla^2 T ({\bf x}) \sigma ({\bf x})\rangle = \int d^2x \, \nabla \cdot ( \nabla  T \sigma ) - \int d^2x \, \nabla T \cdot  \nabla  \sigma $$
Using the Gauss divergence theorem and that $\sigma ({\bf x})$ vanish for $r=R$ it follows that the first term in this equation vanish. For the second term
we use that $\sigma (r)$ only depend on $r=|{\bf x}|$ so that $\nabla \sigma = \sigma'(r) {\bf x}/r$. Thus we get
$$\langle \nabla^2 T ({\bf x}) \sigma ({\bf x})\rangle = -\int_0^R dr \, r \sigma'(r) {\partial \over \partial r} \int_0^{2\pi} d\phi \, T(r,\phi)$$
Using this we can calculate
$${\nabla^2 T_{\rm flash} \over T_{\rm flash}} = {\langle \nabla^2 T ({\bf x}) \sigma ({\bf x})\rangle  \over \langle T ({\bf x}) \sigma ({\bf x}) \rangle} 
\approx -{4.93 \over R^2} \eqno(A6)$$
which is larger than in the static case.

Finally consider
$$\langle \partial_x T ({\bf x}) \sigma({\bf x}) \rangle =
\int d^2x \, \partial_x T \sigma = \int d^2x \, [\partial_x (T \sigma)  -T \partial_x \sigma ]$$
The first term can be written as
$$\int_{-R}^R dy \int_{-h(y)}^{h(y)} dx  \, \partial_x (T \sigma) $$
$$  =  \sigma (R) \int_{-R}^R dy \left [T(h(y),y) -T(-h(y),y) \right ]$$
where $h(y)= (R^2-y^2)^{1/2}$ and where we used that $\sigma(r)$ only depend on $r^2 = h^2(y)+y^2 = R^2$. 
Since $\sigma (R)=0$ the first term vanish.
The second term in can be written as
$$-\int d^2x \,  T \partial_x \sigma = $$
$$- \int_0^R dr \, r \sigma'(r) \int_0^{2\pi} d\phi \, {\rm cos}\phi \, T(r {\rm cos}\phi, r{\rm sin}\phi)$$
Using this equation one obtain
$${1\over d_{\rm slope}} = {T'_{\rm flash} \over T_{\rm flash}} \approx  -{0.734 \over R} \eqno(A7)$$


\begin{thebibliography}{}
\bibitem{Sutter}
G. Sutter and N. Ranc, 
{\it Flash temperature measurement during dry friction
process at high sliding speed},
Wear {\bf 268}, 1237 (2010).

\bibitem{rub1}
B.N.J. Persson,
{\it Rubber friction: role of the flash temperature},
J. Phys.: Condens. Matter {\bf 18}, 7789 (2006).

\bibitem{rub2}
G. Fortunato, V. Ciaravola, A. Furno, B. Lorenz, and B.N.J. Persson, 
{\it General theory of frictional heating with application to rubber friction},
J. Phys.: Condens. Matter {\bf 27}, 175008 (2015).

\bibitem{rub3}
B.N.J. Persson,
{\it Role of Frictional Heating in Rubber Friction},
Tribology Letters {\bf 56}, 77 (2014)



\bibitem{RiceHeat}
J.R. Rice,
{\it Heating and weakening of faults during earthquake slip},
Journal of Geophysical Research: Solid Earth {\bf 111}, 148 (2006).
% http://dx.doi.org/10.1029/2005JB004006

\bibitem{rpp}
S.V. Sukhomlinov, M.H. M\"user and B.N.J. Persson,
{\it Granite sliding on granite: friction, wear rates, surface topography, 
and the scale-dependence of rate-state effects},
Reports on Progress in Physics {\bf 89}, 038301 (2025).

\bibitem{Norway}
S.L. Bore, B.N.J. Persson and H.A. Sveinsson,
{\it Why ice is so slippery},
arXiv preprint arXiv:2603.11539

\bibitem{MarinIce}
A. Atila, S.V. Sukhomlinov, and M.H. M\"user,
{\it Cold Self-Lubrication of Sliding Ice},
Phys. Rev. Lett. {\bf 135}, 066204 (2025).


\bibitem{mechano0}
T.E. Fischer, 
{\it Tribochemsstry},
Ann. Rev. Mater. Sci. {\bf 18}, 303 (1988).

\bibitem{mechano1}
M. Kalin and J. Vizinti,
{\it High temperature phase transformations under fretting conditions},
Wear {\bf 249}, 172 (2001).


\bibitem{MP}
M.H. M\"user and B.N.J. Persson,
{\it On the flash temperature in sliding contacts},
subm. to PRX

\bibitem{flash1}
J.C. Jaeger,
{\it Moving sources of heat and the temperature at sliding contacts},
Proc. Roy. Soc., New South Wales {\bf 56}, 203 (1942).

\bibitem{flash2}
J.F. Archard, 
{\it The temperature of rubbing surfaces}, 
Wear {\bf 2}, 438 (1958).

\bibitem{flash3}
J.A. Greenwood,
{\it An interpolation formula for flash temperatures},
Wear {\bf 150}, {\bf 1-2}, 153 (1991).

\bibitem{flash4}
J.R. Barber,
{\it Distribution of heat between sliding surfaces}, 
J. Mech. Engng. Sci. {\bf 9}, 351 (1967).

\bibitem{flash5}
F.E. Kennedy,
{\it Thermal and thermomechanical effects in dry sliding},
Wear {\bf 100}, 453 (1984).

\bibitem{flash6}
J. Choudhry,  A. Almqvist and R. Larsson,
{\it A Multi-scale Contact Temperature Model for Dry Sliding Rough Surfaces},
Tribology Letters {\bf 69}, 128 (2021).

\bibitem{TianGreen}
X. Tian and F.E. Kennedy,
{\it Maximum and average flash temperatures in sliding contacts},
Journal of Tribology {\bf 116}, 1528 (1994).
% http://dx.doi.org/10.1115/1.2927035

\bibitem{flashAlm2}
J. Choudhry, A. Almqvist and R. Larsson, 
{\it Validation of a Multi-Scale Contact Temperature Model for Dry Sliding Rough Surfaces},
Lubricants {\bf 10}, 41 (2022).

\bibitem{Persson1}
B.N.J. Persson,
{\it On the elastic energy and stress correlation in the contact between elastic solids with randomly rough surfaces},
Journal of Physics: Condensed Matter {\bf 20}, 312001 (2008).

\bibitem{Persson2}
B.N.J. Persson, R. Xu and N. Miyashita
{\it Rubber wear: Experiment and theory},
J. Chem. Phys. 162, 074704 (2025)

\bibitem{Persson3}
N. Miyashita and B.N.J. Persson,
{\it Tire tread block dynamics}, 	
arXiv.2602.22078

\bibitem{size}
M.H. Mi\"user and A. Wang,
{\it  Contact-patch-size distribution and limits of self-affinity in contacts between randomly rough
surfaces},
Lubricants {\bf 6}, 85 (2018).


\bibitem{Mark}
S. Hyun, L. Pei, J.F. Molinari, M.O. Robbins, 
{\it Finite-element analysis of contact between elastic self-affine surfaces},
Phys. Rev. E {\bf 70}, 026117 (2004).

\bibitem{Ruibin}
R. Xu and B.N.J. Persson,
{\it Role of transfer films and interfacial cracking in metallic sliding wear},
subm. to JCP.

\bibitem{Metal1}
A. Tiwari, A. Almqvist and B.N.J. Persson,
{\it Plastic deformation of rough metallic surfaces},
Tribology Letters {\bf 68}, 129 (2020).

\bibitem{Persson}
B.N.J. Persson,
{\it Contact mechanics for randomly rough surfaces},
Surface science reports {\bf 61}, 201 (2006).

\bibitem{plast1}
A. Almqvist and B.N.J. Persson,
{\it Multiscale contact mechanics for elastoplastic contacts},
Physical Review E {\bf 113}, 015503 (2026).

\bibitem{explain}
The smoothing procedure (27) was first used in Ref. \cite{heat1}. It result in an area of real contact with
is virtually identical to the result of the theory for elastoplastic contact for all magnifications.
However, it is not accurate for the describing the separation between the surfaces in the non-contact surface area.
For the interfacial separation another more cumbersome approach has been 
developed \cite{separation}. However, for the flash temperature the contact area is most important 
and it is accurately described using (27).

\bibitem{ReddyhoffHardening}
T. Reddyhoff, A. Schmidt, and H. Spikes,
{\it Thermal Conductivity and Flash Temperature},
Tribology Letters {\bf 67}, 22 (2019).
% http://dx.doi.org/10.1007/s11249-018-1133-8

\bibitem{NATURE}
R. Xu, H. Ren, A. Clerc, G. Mollon, W. Sheng, F. Zhou and B.N.J. Persson,
{\it Sliding contact creates universal self-affine fractal surfaces},
subm to NATURE

\bibitem{heat1}
B.N.J. Persson,  B. Lorenz and A.I. Volokitin,
{\it Heat transfer between elastic solids with randomly rough surfaces},
Eur. Phys. J. E {\bf 31}, 3 (2010). 

\bibitem{separation}
A. Almqvist and  B.N.J. Persson,
{\it Surface separation in elastoplastic contacts},
Physical Review E {\bf 113}, 025509 (2026).

\end{thebibliography}
\end{document}